%% file: icc.tex
\documentclass[conference]{IEEEtran}

\usepackage{graphicx}
\usepackage{amssymb}
\usepackage{amsmath}
\usepackage{epsfig}
\usepackage{cite}
\usepackage{verbatim}
\usepackage{enumerate}
\usepackage{subfigure}
\usepackage{multicol}
\usepackage{pstricks}
\usepackage{pst-node}

\def\blfootnote{\xdef\@thefnmark{}\@footnotetext}

\newtheorem{theorem}{\bf Theorem}

\newtheorem{corollary}{\bf Corollary}

\begin{document}
\title{Capacity Limits of Cognitive Radio with Distributed and Dynamic Spectral Activity}
\author{\authorblockN{Syed Ali Jafar, Sudhir Srinivasa}
\authorblockA{Electrical Engineering and Computer Science\\
University of California Irvine, Irvine, CA 92697-2625\\
Email: {\it syed@uci.edu, sudhirs@uci.edu}	}	}
\maketitle

\IEEEpeerreviewmaketitle

\begin{abstract}
We investigate the capacity of opportunistic communication in the presence of dynamic and distributed spectral activity, i.e. when the time varying spectral holes sensed by the cognitive transmitter are correlated but not identical to those sensed by the cognitive receiver. Using the information theoretic framework of communication with causal and non-causal side information at the transmitter and/or the receiver, we obtain analytical capacity expressions and the corresponding numerical results. We find that cognitive radio communication is robust to dynamic spectral environments even when the communication occurs in bursts of only $3-5$ symbols. The value of handshake overhead is investigated for both lightly loaded and heavily loaded systems. We find that the capacity benefits of overhead information flow from the transmitter to the receiver is negligible while feedback information overhead in the opposite direction significantly improves capacity.
\end{abstract}
\section{Introduction}\label{section:introduction} 
 
Cognitive radio technology has tremendous potential for improving the utilization of radio spectrum. Derived from J. Mitola's doctoral thesis \cite{Mitola_thesis}, a cognitive radio is an intelligent wireless communication system that relies on opportunistic communication between secondary users \footnote{We use the terms \emph{cognitive radio} users and \emph{secondary} users interchangeably.} over temporarily unused spectral bands that are licensed to their primary users \cite{Hillenbrand_etal}. It is driven by software defined radio technology which is in production and available now. However, the development of cognitive radio is still at a conceptual stage due to the multitude of challenges in how the radio learns and adapts to the local spectral activity at each end of the link. Various solutions seek to underlay, overlay or interweave \cite{Fette} the secondary users' signals with the primary users in a way that the primary users of the spectrum are as unaffected as possible \cite{Haykin}. A cognitive radio user may co-exist with the primary users either on a \emph{not-to-interfere} basis or on an \emph{easement} basis \cite{Wazer} which allows secondary transmissions as long as they are below the acceptable interference temperature \cite{Kolodzy}. 



The physical separation of the cognitive radio transmitter and receiver leads to different perspectives on the respective local spectral activity. In general the spectral holes sensed by the transmitter of a cognitive radio  may not be identical to those sensed by the corresponding receiver. One solution around this problem is to have an initial handshake between the transmitter and the receiver in the form of an RTS and CTS (request-to-send and clear-to-send) exchange before the beginning of communication. Such a handshake requires an overhead on the forward link for the RTS signal and on the reverse link for the CTS signal. For the relatively stationary or slowly changing spectral usage scenarios initially envisioned for cognitive radio applications the overhead may be small. However, in order to understand the ultimate performance limits and thus the potential future applications for cognitive radio it is necessary to explore the fundamental limitations on its capacity in a distributed and dynamic spectral environment.


\section{System and Channel Model}\label{sec:SystemModel}
\begin{figure}[h]
\centerline{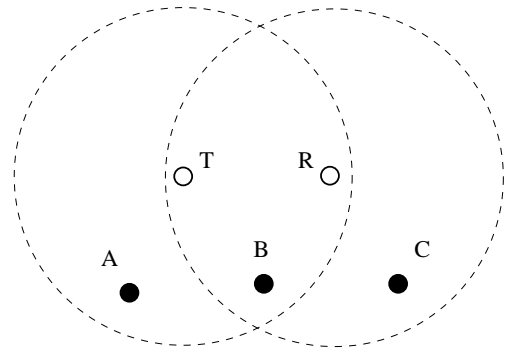}
\caption{Different Perspectives on Local Spectral Activity at Cognitive Radio (Secondary) Transmitter $T$ and Receiver $R$}\label{fig:cog}
\end{figure}

Figure \ref{fig:cog} is a conceptual depiction of a cognitive radio link. The white nodes marked $T$ and $R$ are the cognitive radio transmitter and receiver respectively while the black nodes marked $A, B$ and $C$ are primary users of the spectrum. Given the nature of wireless propagation, spectral activity can be sensed only within a certain locality. The dotted regions around the cognitive radio transmitter and receiver represent the respective \emph{sensing regions}. For the simplified scenario shown in Fig. \ref{fig:cog} cognitive transmitter $T$ can sense when primary users $A$ or $B$ are active and cognitive receiver $R$ can sense when primary users $B$ or $C$ are active. Accordingly, receiver $R$ detects spectral holes when $B$ and $C$ are inactive, while transmitter $T$ detects spectral holes when $A$ and $B$ are inactive. As a consequence, the communication opportunities detected at the transmitter $T$ and receiver $R$ are in general correlated but not identical.

We are interested in how the capacity of cognitive radio is affected by the \emph{distributed} and \emph{dynamic} nature of the spectral environment. These two notions are further explained as follows.
\begin{itemize}
\item {\bf Distributed}: We use the term \emph{distributed} to indicate the different views of local spectral activity at the cognitive transmitter $T$ and receiver $R$ as shown in Fig. \ref{fig:cog}. More precisely we are interested in the correlation of the communication opportunities detected at the transmitter and the receiver. The smaller this correlation, the more \emph{distributed} the system is. A more distributed system has less overlap in the sensing regions of the cognitive transmitter and receiver.
\item {\bf Dynamic}: We use the term \emph{dynamic} to indicate the time variation of the spectral activity of the primary users. Thus, a more dynamic system is one where the spectral activity changes faster and is less predictable. 
\end{itemize}
Intuitively, we expect that as the system becomes more distributed and more dynamic, the capacity will decrease. Our goal in this paper is to make this intuition precise and quantifiable.

The capacity of the system shown in Fig. \ref{fig:cog} depends on a number of variables such as the relative location of the primary and secondary users and the algorithm used to detect the transmission opportunities. As a first step, we start with a basic parametric model for which we compute the capacity. The values of the parameters corresponding to the peculiarities of various physical scenarios can be computed separately in order to match the capacity results to specific system configurations. 

\subsection{Two Switch Channel Model}
\begin{figure}[h]
\centerline{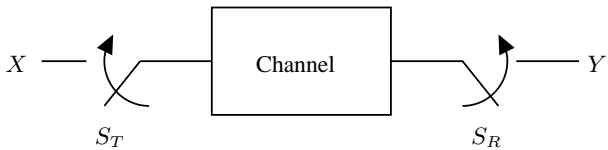}
\caption{Equivalent Channel Model}\label{fig:switch}
\end{figure}

We use the notion of distributed side information to capture the localized spectral activity estimates at the transmitter and the receiver. The system depicted in Fig. \ref{fig:cog} can be reduced to the \emph{switched} channel model shown in Fig. \ref{fig:switch}.
The input $X$ is related to the output $Y$ as
\begin{equation}
Y = (X S_T + N) S_R\label{eq:switch}
\end{equation}
where $N$ is additive white Gaussian noise (AWGN), and $S_T, S_R \in\{0,1\}$ are binary random variables modeled as switches that represent the communication opportunities sensed at the transmitter and receiver. $S_T = 0$, i.e., the transmitter switch is open whenever the transmitter $T$ senses that a primary user is active in its sensing region. Thus the only transmission opportunities for the transmitter correspond to $S_T=1$, when the switch is closed. Similarly, $S_R=0$, or the receiver switch is open when the receiver sees interference from active primary users in its sensing region. Thus, the receiver in this case discards the channel output when it is not perceived to be a communication opportunity. From the receiver's viewpoint, the only communication opportunities correspond to $S_R=1$, when the receive switch is closed.

The key observation here is that \emph{the switch state $S_T$ is known only to the transmitter and the switch state $S_R$ is known only to the receiver}. Therefore, the channel of Fig. \ref{fig:switch} corresponds to communication with distributed side information. Note that the distributed nature of the system is captured in the correlation of $S_T$ and $S_R$ while the dynamic nature is captured in the rate at which the switches change state. We assume that the temporal variation of the switch states follows a block static model. In other words, the switches retain their state for a period of $T_c$ channel uses (one block) after which they change to an independent identically distributed (i.i.d.) state.

The transmit power constraint associated with the channel of Fig. \ref{fig:switch} is 
\begin{equation}\label{eq:power}
\mbox{E}\left[|X|^2S_T\right]\leq P,
\end{equation}
since no power need be transmitted when $S_T=0$.

In general, the channel model may include fading and interference. Moreover the receiver may be on all the time ($S_R=1$). For simplicity of exposition we postpone these additional considerations for later sections and start with the basic model of Fig. \ref{fig:switch} which includes no fading or interference and where the receiver discards the information received when interference from primary user is present. 
\section{Capacity with Dynamic Spectral Activity}\label{sec:dynamic}
The coherence time $T_c$ determines the dynamic nature of the spectral environment. Since single letter capacity expressions are available for memoryless channels, we will start with the assumption that $T_c=1$ i.e., the extreme case of a dynamic environment where each channel use corresponds to an independent spectral usage scenario. Such a scenario could correspond to fast frequency hopping. $T_c=1$ is also an accurate predictor of the capacity limits of cognitive communications where the coding scheme does not utilize the channel memory (i.e. transceivers that are unaffected by interleaving). The case of $T_c>1$ introduces memory into the channel process. In this case, single letter capacity characterizations are not known. However, the channel can still be viewed as a memoryless channel over an extended alphabet ${\mathcal{X}}$ corresponding to $T_c$ channel uses \cite{Marzetta_Hochwald}. With the extended alphabet, the $i^{th}$ channel use corresponds to the transmitted symbol ${\mathcal{X}}(i)=\{X(iT_c), X((iT_c+1), \cdots, X(iT_c+T_c-1)\}$. With the extended alphabet the channel is memoryless and the capacity expressions provided below for $T_c=1$ are directly applicable when we replace the input and output symbols with their corresponding $T_c$ channel use extensions. Both cases of $T_c=1$ and $T_c>1$ are considered in Section \ref{sec:numerical}.

\section{Capacity with Distributed Spectral Activity}\label{sec:distributed}
Capacity with distributed side information at the transmitter and receiver has been studied for the two different cases of causal and non-causal side information at the transmitter.  Unlike transmitter side information, for receiver side information it does not matter whether it is obtained causally or non-causally because the receiver can always wait till the end to decode everything. The corresponding capacity expressions for causal and non-causal side information at the transmitter are presented in this section.

\subsection{Capacity with Non-Causal Side Information - Frequency Coding}
For cognitive radio non-causal side information arises when the coding is performed in the frequency domain. The transmitter and receiver scan the spectral activity in a wideband and communicate opportunistically through a codeword spanning multiple frequency slots that are presumed to be idle.

For the \emph{single user} memoryless channel where i.i.d. side information $S_T$ is available non-causally at the transmitter, the capacity is known to be \cite{Gelfand_Pinsker,Kusnetsov_Tsybakov,Heegard_Gamal, Cover_Chiang, Chiang_Cover}:
\begin{eqnarray}
C^{\mbox{\tiny non-causal}}_{S_T,S_R} = \max_{{\mathcal{P}}_{\mbox{\tiny non-causal}}} I(U;Y,S_R) - I(U;S_T)\label{CnonC}
\end{eqnarray}
where ${\mathcal{P}}_{\mbox{\tiny non-causal}}=\{P(U,X|S_T) =P(U|S_T)P(X|U,S_T)\}$. $U$ is an auxiliary random variable, $X$ and $Y$ are the input and output alphabet, and $S_R$ is i.i.d. side information available at the receiver. 

Comparing this to the case where no side information is available ($S_T=\phi$),
\begin{eqnarray}
C_{\phi,S_R}= \max_{P(U,X)} I(U;Y,S_R) = \max_{P(X)}I(X;Y,S_R), \label{CnoS}
\end{eqnarray}
note that the availability of side information at the transmitter is helpful in that the transmitter can match its input to the channel information by picking the input alphabet $U,X$ conditioned on $S_T$, as opposed to (\ref{CnoS}) where the input can not be matched to the channel state. However, the benefit of matching the input to the channel state comes with the cost of the subtractive term in (\ref{CnonC}), i.e., $I(U;S_T)$ which can be interpreted as the overhead required to inform the receiver about the adaptation to the channel state at the transmitter. 
\subsection{Capacity with Causal Side Information - Temporal Coding}
The case of causal side information represents coding in the time domain. The transmitter and receiver monitor the primary users' activity in a specific frequency slot and opportunistically use it for secondary communications when it is perceived to be idle.

For the case where the side information is available at the transmitter only causally, the capacity expression has been found by Shannon \cite{Shannon1958} as
\begin{equation}
C^{\mbox{\tiny causal}}_{S_T,S_R}  = \max_{P(t)}I(T;Y,S_R)\label{CcS}
\end{equation}
where $T$ is an extended alphabet of mappings from the channel state $S_T$ to the input alphabet $X$.

Recent work has also presented an alternate form for the capacity with causal side information that is directly related to the corresponding expression for the non-causal side information case.
\begin{eqnarray*}
{\begin{array}{lclcl}
C^{\mbox{\tiny non-causal}}_{S_T,S_R} &=& \max_{{\mathcal{P}}_{\mbox{\tiny non-causal}}}&I(U;Y,S_R) - I(U;S_T),  &~\\
C^{\mbox{\tiny causal}}_{S_T,S_R} &=& \max_{{\mathcal{P}}_{\mbox{\tiny causal}}}& I(U;Y,S_R) - I(U;S_T),  &\\
\end{array}}
\end{eqnarray*}
where
\begin{eqnarray*}
{\mathcal{P}}_{\mbox{\tiny non-causal}}&=&\{P(U,X|S_T) = P(U|S)P(X|U,S_T)\}\\
~~~~~{\mathcal{P}}_{\mbox{\tiny causal}}&=&\{P(U,X|S_T) = P(U)P(X|U,S_T)\}
\end{eqnarray*}
In the non-causal case, the choice of $U$ can be made conditional on the channel state $S_T$. In the causal case $U$ is picked independent of $S_T$.  This makes the subtractive term equal to zero for the causal case. In both cases, it suffices for the optimal input symbol $X$ to be just a deterministic function of $U,S_T$.
\section{Capacity of Two Switch Channel Model}
Unlike the direct maximization of mutual information $I(X;Y)$ over the input alphabet distribution $P(X)$ in the absence of side information, capacity characterization in the presence of side information is more involved because of the additional auxiliary random variable $U$. However, as we show in this section, for the two switch channel model of Fig. \ref{fig:switch}, a direct capacity characterization may be obtained for both causal and non-causal side information.
\subsection{Causal Side Information at the Transmitter}
\begin{theorem}\label{theorem:u=x}
For the two switch channel model of Fig. \ref{fig:switch} and causal side information at the transmitter coding can be performed directly on the input alphabet (i.e., $U=X$) and the channel capacity is 
\begin{eqnarray*}
C^{\mbox{\tiny causal}}_{S_T,S_R}(P) & = & \max_{P(X)} I(X;Y,S_R)\\
\end{eqnarray*}
where the channel input $X$ satisfies the power constraint
\begin{eqnarray*}
\mbox{E}\left[|X|^2\right]&=&\frac{P}{\mbox{Prob}(S_T=1)}
\end{eqnarray*}
\end{theorem}
\proof
For causal side information at the transmitter, we know from the previous section that the input $X$ is a deterministic function of the auxiliary random variable $U$ and the transmitter side information, i.e. $X=f(U,S_T)$. Equivalently, let
\begin{eqnarray*}
X&=&\left\{\begin{array}{lr}
              f_1(U), & S_T=1,\\
              f_0(U), & S_T=0.
\end{array}
\right.
\end{eqnarray*}
Then the capacity is unchanged if we instead assume
\begin{eqnarray*}
X&=&f_1(U), ~~S_T=0,1.\label{eq:XfU}
\end{eqnarray*}
The reason is that the channel output is unaffected by $X$ when the switch is open ($S_T=0$). Therefore, the transmitted symbol $X$ is inconsequential for $S_T=0$. Now, according to (\ref{eq:XfU}) $X$ is a deterministic function of the auxiliary random variable $U$ regardless of the channel state. Therefore, decoding $U$ is the same as decoding $X$ and without loss of generality we can replace $U=X$ in the capacity expressions. Furthermore, since $U$ is independent of $S_T$ for causal side information, the input $X$ is also independent of $S_T$.
Therefore, the capacity is the same as when the side information is not available to the transmitter and the power constraint (\ref{eq:power}) can be expressed as
\begin{eqnarray*}
\mbox{E}\left[|X|^2S_T\right]&=&\mbox{E}\left[|X|^2\right]\mbox{E}\left[S_T\right]=P\\
\Rightarrow \mbox{E}\left[|X|^2\right]&=&\frac{P}{\mbox{Prob}(S_T=1)}
\end{eqnarray*}
This completes the proof. \hfill\QED

Theorem \ref{theorem:u=x} leads directly to the following corollary.
\begin{corollary}\label{corollary:causal}
\begin{eqnarray*}
C^{\mbox{\tiny causal}}_{S_T,S_R}(P)=C_{\phi,S_R}\left(\frac{P}{\mbox{Prob}(S_T=1)}\right)
\end{eqnarray*}
The capacity of the two switch channel model with \emph{causal} transmitter side information and transmit power $P$ is the same as the capacity with \emph{no} transmitter side information and transmit power $\frac{P}{\overline S_T}$, where $\overline S_T=\mbox{E}[S_T]$ is the probability that the switch $S_T = 1$, i.e., the average fraction of spectrum that is sensed to be idle at the transmitter.
\end{corollary}
Corollary \ref{corollary:causal} shows that for the two switch channel model of Fig. \ref{fig:switch} there is no benefit of causal side information at the transmitter except the power saving that can be achieved by not transmitting when $S_T=0$. In particular, the optimal codebooks are the same whether the transmitter knows the switch state $S_T$  progressively (causal) or not at all. The only advantage of knowing the switch state is that a higher transmit power codebook may be picked and the power saved by replacing the codeword symbol with $X=0$  whenever $S_T=0$.

Finally, note that although the channel is an additive white Gaussian noise channel, the optimal input distribution is \emph{not} necessarily Gaussian because of the switch $S_T$ whose state is not known to the receiver. Input optimization requires the entropy maximization of a mixture process, consisting of the AWGN and the transmitted symbol $X$. Gaussian mixture models are traditionally used for classification in pattern recognition literature based on the maximum entropy principle. Therefore, we will use a Gaussian input distribution for our capacity calculations. Note that Gaussian inputs may not be strictly optimal, but they do represent an innerbound that we expect to be fairly tight based on the corresponding results in entropy maximization with Gaussian mixture processes \cite{Keysers_Och_Ney}.
\subsection{Non-Causal Side Information at the Transmitter}
\begin{theorem}\label{theorem:nc}
For the two switch channel model of Fig. 2 and non-causal side information at the transmitter, coding can be performed directly on the input alphabet (i.e., $U=X$) and the channel capacity is 
\begin{eqnarray*}
C^{\mbox{\tiny non-causal}}_{S_T,S_R}(P) & = & \max_{P(X|S_T)} I(X;Y,S_R)-I(X;S_T)
\end{eqnarray*}
with the power constraint $\mbox{E}[|X|^2S_T]=P$.
\end{theorem}
The proof proceeds as in the causal case so we have $X=f_1(U), S_T=0,1$ and therefore we can set $X=U$. However, there is a difference. In the non-causal case, $U$ depends on $S_T$ and therefore, the input $X$ is not independent of the transmitter side information. Interestingly, even though the output is not affected by the transmitted symbol when $S_T=0$, the choice of the input distribution $P(X|S_T=0)$ does affect the mutual information $I(X;Y,S_R)$. Therefore, one cannot arbitrarily set $P(X|S_T=0)=P(X|S_T=1)$ as in the causal case. Another observation that can be made is the analogy between the current problem and the memory with stuck-at defects explored in \cite{Kusnetsov_Tsybakov,Heegard_Gamal}. The open switch state is similar to the stuck-at defect because when $S_T=0$ the output is independent of the input, i.e. $P(Y|X,S_T=0)=P(Y|S_T=0)$.

Based on the similarity of our two switch channel model and the memory with stuck-at defects explored in \cite{Heegard_Gamal} and the optimal input distribution presented in \cite{Heegard_Gamal} we pick the input distribution as:
\begin{eqnarray*}
X = {\mathcal{N}}(0,P),&& S_T=1 \\
X= {\mathcal {N}}(0,\alpha), &&S_T=0.
\end{eqnarray*}
and we optimize over $\alpha$. Again, this choice of input distribution is an innerbound on capacity.

Next we present several capacity outerbounds which complement the results presented above as well as represent the precise capacity under some additional assumptions.
\section{Capacity Outerbounds}
\subsection{Capacity with Global Side Information: $C_{\star,\star}$}
Global side information refers to the case where both $S_T$ and $S_R$ are known to both the transmitter and the receiver. 
In this case the transmitter and the receiver know exactly which spectral holes are unused at both ends of the cognitive radio link. Therefore, the transmission is restricted to the slots where $S_T=1,S_R=1$. Gaussian inputs are optimal in this case and the capacity expression is:
\begin{eqnarray*}
C_{\star,\star}(P)=\mbox{Prob}(S_TS_R=1)\log\left(1+\frac{P}{\mbox{Prob}(S_TS_R=1)}\right)
\end{eqnarray*}
Notice that in this case, capacity with causal side information is identical to capacity with non-causal side information.
\subsection{Capacity with Full Side Information at Receiver: $C_{S_T,\star}$}
Full side information at the receiver refers to the case where both $S_T$ and $S_R$ are known to the receiver. The transmitter is still assumed to know only $S_T$. In previous work \cite{Syed_allerton} we have shown that when the transmitter side information is also available at the receiver, the capacity is the same with causal or non-causal side information at the transmitter. In this case also, Gaussian inputs are optimal and the capacity is given by:
\begin{eqnarray*}
C_{S_T,\star}(P)=\mbox{Prob}(S_TS_R=1)\log\left(1+\frac{P}{\mbox{Prob}(S_T=1)}\right)
\end{eqnarray*}
\subsection{Capacity with Full Side Information at Transmitter: $C_{\star,S_R}$}
Full side information at the transmitter refers to the case where both $S_T$ and $S_R$ are known to the transmitter. The receiver is still assumed to know only $S_T$. This is of practical significance since it represents the capacity with a channel state feedback channel from the receiver to the transmitter. Through this feedback channel the transmitter can potentially learn the instantaneous channel state at the receiver and adapt its transmit strategy accordingly. The following relationship is easily obtained for the \emph{causal} case:
\begin{eqnarray*}
C_{\star,S_R}^{\mbox{\tiny causal}}(P)=C_{S_T,S_R}^{\mbox{\tiny causal}}\left(\frac{P}{\mbox{Prob}(S_R=1|S_T=1)}\right)
\end{eqnarray*}
The corresponding expression for the non-causal case is
\begin{eqnarray*}
C^{\mbox{\tiny non-causal}}_{\star,S_R}(P) & = & \max_{P(X|S_T,S_R)} I(X;Y|S_R)-I(X;S_T|S_R)
\end{eqnarray*}
with the power constraint $\mbox{E}[|X|^2S_T]=P$.

\section{Numerical Results}\label{sec:numerical}
\begin{figure}[t]
\includegraphics[height=3.4in,width=9cm]{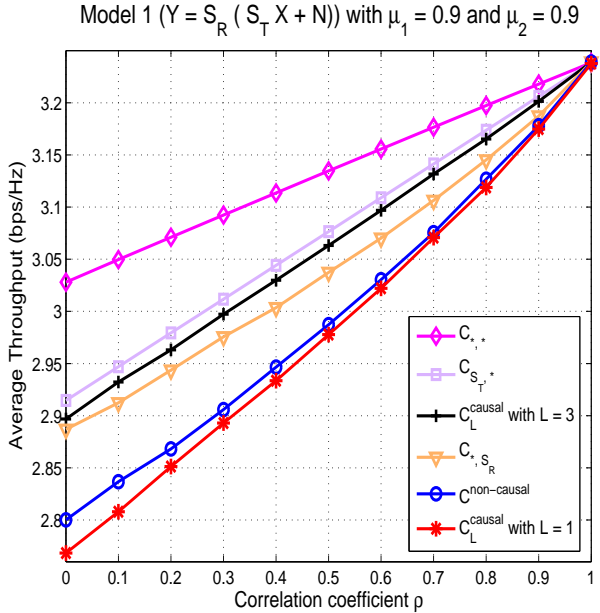}
\caption{Capacity of the two switch model in a lightly loaded system}\label{fig:underpopulated}
\end{figure}
\begin{figure}[t]
\includegraphics[height=3.4in,width=9cm]{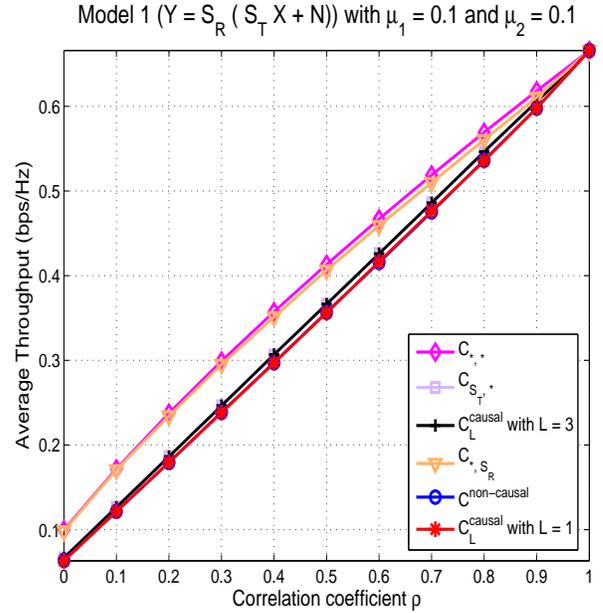}
\caption{Capacity of the two switch model in a heavily loaded system}\label{fig:overpopulated}
\end{figure}
To gain numerical insights we compute the capacity expressions presented in the preceding sections. Figures \ref{fig:underpopulated} and \ref{fig:overpopulated} represent cognitive communication scenarios where, on the average, active primary users occupy 10\% (lightly loaded system) and 90\% (heavily loaded system) of the available spectrum, respectively. Thus, for the basic model of (\ref{eq:switch}) the lightly loaded and heavily loaded scenarios correspond to $\mbox{E}[S_T]=\mbox{E}[S_R]=0.9$ and $0.1$, respectively. For transmit power of $10$ dB, The average throughput is plotted as a function of the correlation  coefficient $\rho$ between $S_T$ and $S_R$. As mentioned before, $\rho$ quantifies the distributed nature of the spectral activity. 

Several interesting observations can be made from these plots. First, unlike the case of the memory with stuck-at defects considered in \cite{Heegard_Gamal}, non-causal side information does not appear to offer a significant advantage over causal side information for the input distributions considered. This advantage disappears even further for an overpopulated system. Second, while $C_{S_T,\star}$ dominates $C_{\star,S_R}$ in an underpopulated system, the opposite is true in an overpopulated system. This suggests that transmitter side information is more valuable in an overpopulated system while receiver side information is more valuable in an underpopulated system. Thirdly, the plots reveal how the dynamic nature of the spectral activity affects the capacity. Intuitively one would expect that as the channel coherence time increases the receiver will acquire more knowledge of the transmitter state $S_T$. Mathematically, we expect that as $T_c\rightarrow\infty$, $C^{\tiny causal}_{S_T,S_R}(P)\rightarrow C^{\tiny causal}_{S_T,\star}(P)$. However, the plots indicate that the convergence is very rapid. Even with a channel coherence time as small as $3$ channel symbols, the capacity $C^{\tiny causal}_{S_T,S_R}(P)$ is approximately the same as with perfect knowledge of the communication opportunities at the receiver. 

The rapid convergence of $C^{\tiny causal}_{S_T,S_R}(P)$ to $C^{\tiny causal}_{S_T,\star}(P)$ is a positive indicator of the capabilities of cognitive radio systems under highly dynamic spectral environments. Cognitive radio is based on the premise of minimal interference to the primary users of the spectrum. In order to ensure this, the cognitive communications must occur over short bursts, allowing the secondary users to frequently check for primary users' activity and to jump out of the active bands as soon as the primary users' presence is detected. If the primary users' spectral activity is highly dynamic, the length of the opportunistic communication bursts must be very small to ensure minimal interference.  The numerical results show that even with bursts as small as $3-5$ channel uses, the capacity achieved is as high as with full knowledge of the spectral holes at the receiver. The numerical results indicate that cognitive communication is fairly robust to the dynamic nature of the spectral environment. 

Finally, note that while the capacity benefits of sharing the transmitter side information with the receiver are automatically obtained for $T_c > 3$, the additional benefits of sharing the receiver side information with the transmitter are significant and not automatically obtained. In other words, overhead information  from the receiver to the transmitter improves capacity significantly but the overhead of information flow from the transmitter to the receiver presents no benefit when communication bursts last longer than $3$ channel symbols. 
\section{Extensions}
For simplicity, we presented our results for the two switch model of Fig. \ref{fig:switch}.  The analysis as well as the numerical insights of the previous sections extend to more complex models when we incorporate additional considerations such as fading, interference and channel knowledge. For want of space, we will briefly summarize some of these extensions and leave a detailed exposition to \cite{Syed_Full}. 

For a general model we include channel fading and interference. We also allow the receiver to process all received signals. Notice that unlike the transmitter which needs to shut off transmission to avoid interference to the primary users, the receiver can stay on all the time. While this can increase the signal processing required at the receiver, it represents a more capable system with a higher capacity. Such a system can be represented by the following system model:
\begin{equation}
Y = S_T X + S_R N, \label{eq:model2}
\end{equation}
where, $Y, X, S_T$ and $N$ are defined as before. However, instead of a binary switch, $S_R$ represents a general channel state. Notice that both the channel fade and the interference (assumed Gaussian) power can be included into the AWGN normalization factor $S_R$. A high value of $S_R$ represents poor communication conditions, either because the channel is severely faded or because the interference from the active users near the receiver is too strong. Note that we do not allow the cognitive radio receiver to decode and subtract the interference. As before, the assumption is that $S_T$ is known only to the transmitter and $S_R$ is known only to the receiver. 
\begin{figure}[t]
\includegraphics[height=3.02in,width=9cm]{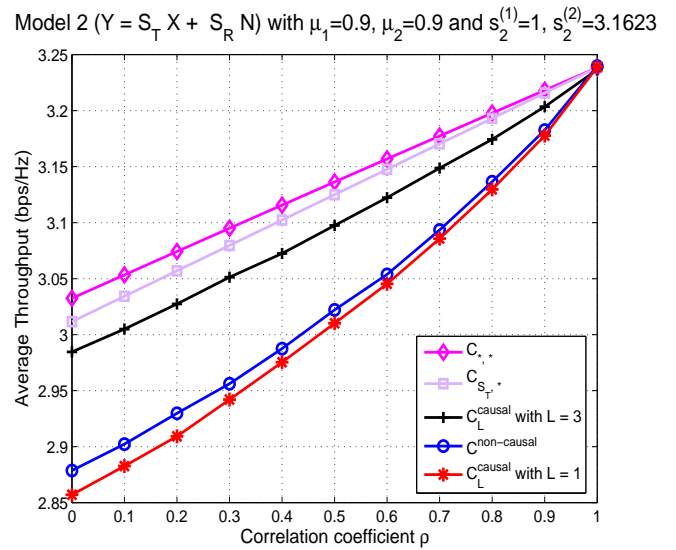}
\caption{Capacity of the general model in a lightly loaded system}\label{fig:general}
\end{figure}

The numerical results for the general model are shown in Fig. \ref{fig:general}. For the plot we have assumed a two state channel with a good and bad state corresponding to SNR of $10$ dB and $0$ dB respectively. If we choose bad channel as $-10$ dB, the performance is identical to the two switch model.  This shows that the receiver can ignore the received signal in bad SNR conditions without a significant capacity loss. More elaborate fading models lead to similar results and are included in \cite{Syed_Full}.
\section{Conclusion}
We explore the capacity of a cognitive radio system when the communication opportunities detected at the secondary transmitter and receiver are correlated but not identical. The problem is formulated as communication with distributed and time varying side information at the transmitter and receiver. We develop a two-switch model for which we present capacity expressions as well as inner and outer bounds. Using the capacity expressions we determine the value of side information and the necessity of the overhead associated with a handshake between the transmitter and receiver to initiate opportunistic communication. 
\label{conclusion}

\bibliographystyle{ieeetr}
\bibliography{Thesis}

\end{document}

%% file: cog.pstex_t
\begin{picture}(0,0)%
\includegraphics{cog}%
\end{picture}%
\setlength{\unitlength}{2960sp}%
\begingroup\makeatletter\ifx\SetFigFont\undefined%
\gdef\SetFigFont#1#2#3#4#5{%
  \reset@font\fontsize{#1}{#2pt}%
  \fontfamily{#3}\fontseries{#4}\fontshape{#5}%
  \selectfont}%
\fi\endgroup%
\begin{picture}(4112,2857)(303,-3070)
\put(1051,-2386){\makebox(0,0)[lb]{\smash{\SetFigFont{9}{10.8}{\rmdefault}{\mddefault}{\updefault}\special{ps: gsave 0 0 0 setrgbcolor}A\special{ps: grestore}}}}
\put(2326,-2311){\makebox(0,0)[lb]{\smash{\SetFigFont{9}{10.8}{\rmdefault}{\mddefault}{\updefault}\special{ps: gsave 0 0 0 setrgbcolor}B\special{ps: grestore}}}}
\put(1876,-1561){\makebox(0,0)[lb]{\smash{\SetFigFont{9}{10.8}{\rmdefault}{\mddefault}{\updefault}\special{ps: gsave 0 0 0 setrgbcolor}T\special{ps: grestore}}}}
\put(2701,-1561){\makebox(0,0)[lb]{\smash{\SetFigFont{9}{10.8}{\rmdefault}{\mddefault}{\updefault}\special{ps: gsave 0 0 0 setrgbcolor}R\special{ps: grestore}}}}
\put(3676,-2311){\makebox(0,0)[lb]{\smash{\SetFigFont{9}{10.8}{\rmdefault}{\mddefault}{\updefault}\special{ps: gsave 0 0 0 setrgbcolor}C\special{ps: grestore}}}}
\end{picture}

%% file: switch.pstex_t
\begin{picture}(0,0)%
\includegraphics{switch}%
\end{picture}%
\setlength{\unitlength}{2960sp}%
\begingroup\makeatletter\ifx\SetFigFont\undefined%
\gdef\SetFigFont#1#2#3#4#5{%
  \reset@font\fontsize{#1}{#2pt}%
  \fontfamily{#3}\fontseries{#4}\fontshape{#5}%
  \selectfont}%
\fi\endgroup%
\begin{picture}(4875,1147)(826,-1186)
\put(826,-586){\makebox(0,0)[lb]{\smash{\SetFigFont{9}{10.8}{\rmdefault}{\mddefault}{\updefault}\special{ps: gsave 0 0 0 setrgbcolor}$X$\special{ps: grestore}}}}
\put(5701,-586){\makebox(0,0)[lb]{\smash{\SetFigFont{9}{10.8}{\rmdefault}{\mddefault}{\updefault}\special{ps: gsave 0 0 0 setrgbcolor}$Y$\special{ps: grestore}}}}
\put(2926,-586){\makebox(0,0)[lb]{\smash{\SetFigFont{9}{10.8}{\rmdefault}{\mddefault}{\updefault}\special{ps: gsave 0 0 0 setrgbcolor}Channel\special{ps: grestore}}}}
\put(1576,-1186){\makebox(0,0)[lb]{\smash{\SetFigFont{9}{10.8}{\rmdefault}{\mddefault}{\updefault}\special{ps: gsave 0 0 0 setrgbcolor}$S_T$\special{ps: grestore}}}}
\put(4726,-1186){\makebox(0,0)[lb]{\smash{\SetFigFont{9}{10.8}{\rmdefault}{\mddefault}{\updefault}\special{ps: gsave 0 0 0 setrgbcolor}$S_R$\special{ps: grestore}}}}
\end{picture}